\documentclass[fleqn,12pt,twoside]{article}
\usepackage{espcrc1}

\usepackage{graphicx}
\usepackage[figuresright]{rotating}

\newcommand{\AmS}{{\protect\the\textfont2
  A\kern-.1667em\lower.5ex\hbox{M}\kern-.125emS}}

\newcommand{\be}{\begin{eqnarray}}
\newcommand{\ee}{\end{eqnarray}}

\title{Scattering through QCD Sphalerons}

\author{I. Zahed
 \thanks{Supported in part by US DOE grant DE-FG-88ER40388.}\\
Department of Physics and Astronomy,\\ State University of New York,
Stony Brook NY 11794-3800
                }
       
\begin{document}

\maketitle

\begin{abstract}
Diffractive parton-parton scattering in the soft pomeron regime 
can be explained by the production of a QCD sphaleron. Sphaleron
production results into the emission of $3+2N_F$ gluons and quarks.
At RHIC we expect many sphalerons to be released thereby affecting
most prompt processes.
\end{abstract}

\section{Introduction}

In perturbative processes gluon production is coupling constant 
suppressed. In instanton induced processes gluon production is
suppressed by the smallness of the tunneling parameter $\kappa$.
Diffractive parton-parton scattering at large $\sqrt{s}\gg\sqrt{-t}$ 
receives a substantial contribution from instantons~\cite{sz01,KKL}. 
The soft pomeron intercept is proportional to $\kappa\approx 10^{-2}$ 
giving a simple explanation to its smallness. Instanton-induced
diffractive scattering does not account for the odderon. 

A semiclassical description of parton-parton diffractive scattering
in QCD can be achieved using singular gauge configurations (Landau
method). The singularities are essential
in interpolating between a vacuum configuration with zero
energy and the escape configuration with finite energy $Q$~\cite{khlebnikov}. 
In the semiclassical approximation, the cross section rises
due to the initial increase in the gluon phase space, and falls
due to the lack of overlape between the initial and final state. 
The maximum takes place at the sphaleron mass $M_S=3\pi/4\alpha\rho$ 
for a fixed size $\rho$. Typically $\rho\approx 1/3$ fm with 
$M_S\approx 3$ GeV.

\section{Singular Yang-Mills}

For large energies $Q\approx 1/T$ or small Euclidean times 
$T/\rho\ll 1$, the singular gauge configurations simplify.
For $O(3)$ symmetry  in the Euclidean time interval
$|t|\leq T/2$, the gauge configuration is (temporal gauge)

\be
A_i^a (\vec{x}, t) \approx 
(\delta_{ai}-n_an_i)\,D(r,t)/2r + n_an_i\,D(r,t)/2r
\label{S8}
\ee
where $n_i=\vec{x}_i/r$ is a unit 3-vector, and $|\vec{x}|=r\geq 0$
a radial variable.  $D(r,t)$ is continuous and differentiable everywhere, 
except at $r=0$ for $t=\pm T/2$ where it is singular, i.e.
$D(r, \pm T/2) = -{4\rho}/r$. At the escape point $t=0$, the gauge
configuration reads

\be
D(r, 0) \approx  \frac {4\rho \,r}{r^2+ 0.763\,\rho\,T}\,\,
\label{XS5}
\ee
with radial energy density ($\lambda=(Q/M_S)^{1/5}$)

\be
\Theta_{00} (r,0) \approx  \frac {4\pi}{g^2} \,
\frac{24\,\rho^4\,r^2}{(r^2+ \lambda^2 \rho^2)^4}\,\,,
\label{XS6}
\ee
that integrates to $Q$. The tunneling time  $T$ relates to 
the energy $Q$ through $T\approx 1.3/\lambda^2\rho$. 
The Chern-Simons number at the turning point is ${\bf N}=\lambda^2/2$.
For $Q>M_S$ the initial configuration follows from the sphaleron by
a simple rescaling of the size $\rho\rightarrow \rho/\lambda$ and the
energy density $\Theta_{00}\rightarrow \lambda^4\,\Theta_{00}$.

\section{Outgoing Gluons}

At the escape point, the gauge configuration (\ref{S8}-\ref{XS5})
undergoes an explosive evolution in Minkowski space~\cite{jsz02}.
The large time asymptotic of the transverse gluon field in Lorentz
gauge admits a normal mode decomposition

\be
{\cal A}_i^a (t, \vec{k}) =\frac{(2\pi)^{{3}/{2}}}{\sqrt{2k}}\, 
\left(\lambda_i^m (\vec{k})\,b^{am} (\vec{k}) e^{-ikt} + 
\lambda_i^m (-\vec{k})\,b^{am\,*} (-\vec{k}) e^{+ikt}\right)
\label{LS6}
\ee
with 2 real polarizations,

\be
\lambda_i^m (\vec{k})\,b^{am} (\vec{k}) = \frac{\rho}{g\sqrt{\pi\,k}}\,
(-i\epsilon_{aij}\hat{k}_j \,J(k\rho) + (\delta_{ai}-\hat{k}_a\hat{k}_i)
\,J'(k\rho))\,\,,
\label{LS7}
\ee
and

\be
J(k\rho) = 2\,{\rm Re}\,\int_0^{\pi/2}\,dy\,e^{ik\rho\,{\rm cotan}\,y}\,
\left(1-\frac{\sqrt{2}}{{\rm cosh}(\sqrt{2}\,y)}\right)\,\,,
\label{LS8}
\ee
at the sphaleron point. The ensuing gluon multiplicity for all
$\lambda$'s is

\be
{\bf n} (k) \sim \lambda^4\,\frac{8k\rho^2}{g^2\lambda^2}\,
\left( J^2 \left(\frac{k\rho}{\lambda}\right) + 
{J'}^2\left(\frac{k\rho}{\lambda}\right) \right)\,\,.
\label{LS10x}
\ee
and is displayed in Fig.~1 for  $\lambda=0.5,1,2$. At the sphaleron
point, it integrates to about 3 gluons.


\begin{figure}[ht!]
\begin{center}
      \includegraphics[width=2.in]{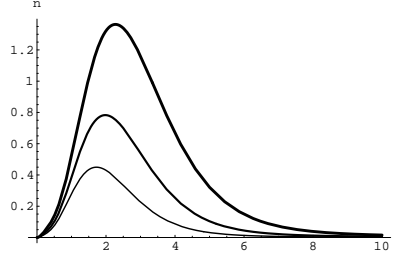}
\end{center}
\vskip -0.4cm
  \caption[]
  {Density of emitted gluons for $\lambda=0.5,1,2$.}
\end{figure}


\section{Outgoing Quarks}

The solution of the Dirac equation in the explosive sphaleron background
in Minkowski space is involved. However, the $O(3)$ zero energy states can be
obtained exactly by conformal mapping~\cite{sz02}. The result is

\be
{\bf Q}_+ (t,r) = 
\frac{{\bf C}}{((t+i\rho)^2-r^2)^2}\,
\left((\rho-it) +i\vec{\sigma}\cdot\vec{x}\right)\,e^{-2F(\xi)}\,U_+\,\,,
\label{Z19}
\ee
with $F(\xi)$ a smooth complex function. Fig.~2 shows the time evolution
of the quark density in space after the explosive production.


\begin{figure}[ht!]
\begin{center}
      \includegraphics[width=2.in]{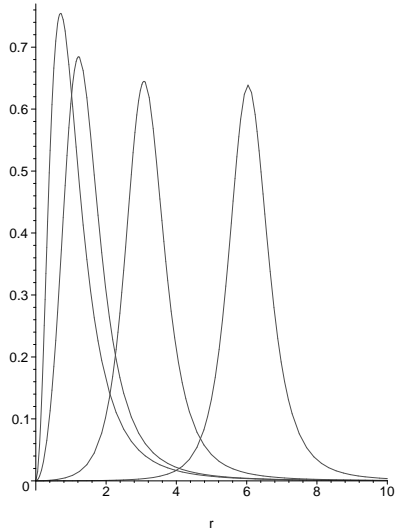}
\end{center}
\vskip -0.4cm
  \caption[]
  {Quark scalar density at different times.}
\end{figure}


A normal mode decomposition allows for a spectral analysis of the final
quark states released by the sphaleron. Using the free
field decomposition asymptotically,

\be
&&{\bf Q}_+ (t,k) = \frac{(2\pi)^{\frac 32}}{\sqrt{2k}}
\,\left({\bf q}_R (\vec{k}) e^{-ikt} + {\bf q}_L^\dagger (-\vec k ) \,e^{+ikt}\right)\,\,,
\label{Z25}
\ee
where ${\bf q}_{R,L}$ refer to right-left handed chiral quarks,

\be
{\bf q}_{R,L}^\dagger (k) = 2\rho^3\,\sqrt{\frac k{\pi}}\,e^{-k\rho}\,
(1\pm\vec{\sigma}_s\cdot\hat{k})\,U_\mp\,\,.
\label{Z25xx}
\ee
The chiral density of left antiquarks and right quarks
are opposite asymptotically 

\be
{\bf n}_R (k) = -{\bf n}_L (k) = \frac {4\pi\,k^2}{2k}
 \,|{\bf q}_R^\dagger(\vec k )|^2 = \rho\,(2k\rho)^2\,e^{-2k\,\rho}\,\,.
\label{Z26}
\ee
The distribution integrates exactly to one produced quark,
in agreement with general spectral flow analysis.
The quark spectrum is close to Planckian with an effective
temperature $T=2/\rho$ of about 300 MeV for a standard $\rho=1/3$ fm.
The released quarks carry asymptotically a total energy ${\bf
M}_F=3/\rho$, which is small in the weak coupling limit, i.e.
${\bf M}_F/{\bf M}_S = \alpha/\pi\ll 1$.

\section{Conclusions}

Diffractive parton-parton scattering through QCD sphalerons produces
about 3 gluons and 2$N_F$ light quarks. The production is explosive 
and occurs on a natural tunneling time scale of order $\rho\sim 1/3$ fm,
with a typical energy release of about 3 GeV. 
In pp scattering we expect about 1 sphaleron to be produced, while in NN
scattering in the RHIC regime we expect hundreds of sphalerons to
be released. Clearly these strongly coherent chromomagnetic
fields should affect most of the prompt processes at RHIC (see
E. Shuryak talk in this meeting).



\begin{thebibliography}{99}
\bibitem{sz01}
E.~Shuryak and I.~Zahed, Phys. Rev. {\bf D62} (2000) 085014; 
M.~Nowak, E.~Shuryak and I.~Zahed, Phys. Rev. {\bf D64} (2001) 034008.

\bibitem{KKL}
D.~Kharzeev,~Y.~Kovchegov~and~E.~Levin, Nucl. Phys. {\bf A690} (2001) 621.


\bibitem{khlebnikov}
S.~Khlebnikov, Phys. Lett. {\bf B282} (1992) 459;
D.~Diakonov and V.~Petrov, Phys. Rev. {\bf D 50} (1994) 266.


\bibitem{jsz02}
R.~Janik, E.~Shuryak and I.~Zahed, {\tt hep-ph/0206005}.

\bibitem{sz02}
E.~Shuryak and I.~Zahed, {\tt hep-ph/0206022}.


\end{thebibliography}
\end{document}